%% file: main.tex
\documentclass{article}
\usepackage[T1]{fontenc}
\usepackage{spconf,amsmath,amssymb,graphicx,booktabs,multirow,url}
\usepackage{tikz}
\usetikzlibrary{arrows.meta,positioning,calc}
\usepackage[hidelinks]{hyperref}
\usepackage{etoolbox,needspace,balance}
\apptocmd{\thebibliography}{\setlength{\itemsep}{0pt}\setlength{\parskip}{0pt}}{}{}
\graphicspath{{figs/}}

\renewcommand{\topfraction}{0.95}\renewcommand{\dbltopfraction}{0.95}
\renewcommand{\textfraction}{0.05}\renewcommand{\floatpagefraction}{0.8}
\usepackage{eso-pic}
\AddToShipoutPictureFG*{%
  \put(0,34){\makebox[\paperwidth]{\parbox{6.4in}{\centering\footnotesize
    This work has been submitted to the IEEE ICASSP for possible publication.
    Copyright may be transferred without notice, after which this version
    may no longer be accessible.}}}}

\newcommand{\proxy}{ProxyMOS}
\newcommand{\mosds}{mos260}
\newcommand{\best}[1]{\textbf{#1}}

\newcommand{\link}[1]{\href{https://#1}{\nolinkurl{#1}}}

\title{ProxyMOS: Label-Free Speech Quality Assessment by\\Multi-Teacher Distillation with Adaptive Routing}
\name{Maxim Trokunov$^{1}$ \qquad Kirill Borodin$^{1,2}$ \qquad Nikita Vasiliev$^{1,2}$ \qquad Grach Mkrtchian$^{2,3}$}
\address{$^{1}$Lab260, Yerevan, Armenia \qquad $^{2}$BitmanagerAI, Dubai, UAE \qquad $^{3}$MTUCI, Moscow, Russia\\
kborodin.research@gmail.com}

\begin{document}
\ninept
\renewcommand{\footnotesize}{\fontsize{9}{10.5}\selectfont}  
\raggedbottom  
\maketitle

\begin{abstract}
Human mean opinion scores (MOS) are costly to collect, and non-intrusive MOS predictors degrade sharply outside their training domain.
\proxy{} turns a pool of public MOS predictors into a single stronger model without new human labels.
Eight predictors are benchmarked against human ratings; the five most informative enter a subset search under uniform, correlation-weighted, error-weighted, MSE-optimised and adaptive per-utterance routing; and the best routed four-model ensemble labels 807k unlabeled utterances that train a wav2vec~2.0 student.
On URGENT the student reaches Spearman $\rho=0.802$ against $0.773$ for the best teacher.
On \mosds{}, a new Russian TTS benchmark of 4{,}600 utterances from 38 synthesis conditions, it reaches $\rho=0.636$ against $0.613$ per utterance and $0.95$ per condition, matching its own routed ensemble in one forward pass.
Adaptive routing is the only rule that does not degrade when weak predictors are added.
Model, ONNX exports and \mosds{} are released.
\end{abstract}


\section{Introduction}
\label{sec:intro}
Perceptual quality drives the value of text-to-speech (TTS), voice conversion, codecs and speech enhancement, and the community measures it with the mean opinion score (MOS) defined in ITU-T P.800~\cite{itu_p800}.
Listening tests are slow, costly and noisy~\cite{lemaguer2024limits}, so non-intrusive neural predictors have become the practical substitute for intrusive metrics such as PESQ.
MOSNet~\cite{mosnet} regressed MOS from spectrograms; NISQA~\cite{nisqa} and DNSMOS~\cite{dnsmos} target telephony and enhancement; and self-supervised front-ends such as HuBERT, XLS-R and Whisper underpin SSL-MOS~\cite{sslmos}, UTMOS~\cite{utmosv2}, Distill-MOS~\cite{distillmos} and WhiSQA~\cite{whisqa}.

Two problems persist.
First, every predictor is tied to the domain of its training labels: the VoiceMOS challenges~\cite{voicemos2022} and MOS-Bench report large drops under domain shift, and we observe rank correlations that fall from $0.77$ to $0.47$ for the same model between enhanced natural speech and Russian TTS.
Second, closing that gap with more human labels is exactly the cost the predictors were meant to avoid.

Ensembles are the standard remedy for the first problem and knowledge distillation~\cite{hinton2015distill} for the cost of deploying them, and both have precedents in MOS prediction: challenge systems stack SSL predictors with a learned combiner~\cite{utmos,voicemos2022}, RAMP weights predictors per utterance by confidence~\cite{ramp}, MTQ-Net trains on pseudo-labels from a pretrained assessor~\cite{mtqnet}, VoiceMOS 2024 added a semi-supervised track~\cite{voicemos2024}, and Distill-MOS distils one XLS-R-SQA teacher into a compact student on unlabeled speech~\cite{distillmos}.
Less studied is their combination: heterogeneous public predictors, a per-utterance combiner learned on existing labels, and distillation of the combined signal into one model.
We treat public MOS predictors as noisy annotators, learn how to combine them on the small labeled sets that already exist, and then let the combination annotate large unlabeled corpora at no cost.
A single student distilled from these targets inherits the ensemble's robustness while keeping single-model inference.
Our contributions are:
\begin{itemize}\setlength{\itemsep}{0pt}
\item A cross-domain benchmark of eight public MOS predictors on enhanced natural speech (URGENT) and synthetic Russian speech (\mosds{}), with 95\% confidence intervals, showing that no single predictor is best in both domains (Sec.~\ref{sec:teachers}).
\item A search over 26 teacher subsets and five aggregation rules. On URGENT, uniform averaging of the right four models already beats every teacher; on the held-out benchmark only a per-utterance router survives weak members (Sec.~\ref{sec:ensembles}).
\item \proxy{}\footnote{Model: \link{huggingface.co/lab260/ProxyMos}; benchmark: \link{huggingface.co/datasets/lab260/mos260}; code and ONNX exports: \link{github.com/lab260ru/ProxyMOS}.}, a wav2vec~2.0 student distilled from 807k ensemble-labeled utterances that outperforms all its teachers on both benchmarks and matches its own teacher ensemble on the held-out domain (Sec.~\ref{sec:student}).
\item \mosds{}, a 4{,}600-utterance Russian TTS MOS benchmark spanning codecs, vocoders, zero-shot and diffusion TTS, with 11.5k quality and 28k intonation ratings from native listeners after reliability filtering (Sec.~\ref{sec:mos260}).
\end{itemize}

\section{Method}
\label{sec:method}
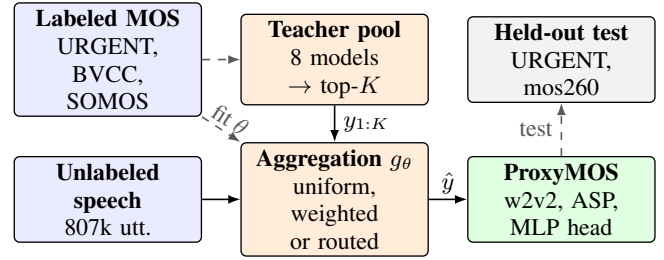
\begin{figure}[t]
\centering
\begin{tikzpicture}[x=1cm,y=1cm,font=\small,
  box/.style={draw,rounded corners=2pt,align=center,inner sep=2.5pt,line width=0.5pt,minimum height=0.85cm,text width=2.3cm},
  arr/.style={-{Latex[length=1.6mm,width=1.2mm]},line width=0.6pt},
  darr/.style={arr,dashed,gray!70!black},
  lab/.style={font=\small,inner sep=1pt}]
\node[box,fill=blue!8]    (lab) at (1.3,2.3) {\textbf{Labeled MOS}\\URGENT, BVCC,\\SOMOS};
\node[box,fill=blue!8]    (unl) at (1.3,0.45) {\textbf{Unlabeled}\\\textbf{speech}\\807k utt.};
\node[box,fill=orange!15] (tea) at (4.3,2.3) {\textbf{Teacher pool}\\8 models\\$\to$ top-$K$};
\node[box,fill=orange!15] (agg) at (4.3,0.45) {\textbf{Aggregation} $g_\theta$\\uniform, weighted\\or routed};
\node[box,fill=green!12]  (stu) at (7.3,0.45) {\textbf{\proxy{}}\\w2v2, ASP,\\MLP head};
\node[box,fill=gray!10]   (ev)  at (7.3,2.3) {\textbf{Held-out test}\\URGENT,\\\mosds{}};
\draw[darr] (lab) -- (tea);
\draw[darr] (lab.south east) -- node[lab,sloped,above=0.5pt,pos=0.55]{fit $\theta$} (agg.north west);
\draw[arr]  (unl) -- (agg);
\draw[arr]  (tea) -- node[lab,right=2pt]{$y_{1:K}$} (agg);
\draw[arr]  (agg) -- node[lab,above=1pt]{$\hat y$} (stu);
\draw[darr] (stu) -- node[lab,left=2pt]{test} (ev);
\end{tikzpicture}
\vspace{-4pt}
\caption{Pipeline. Labeled sets rank the teachers (dashed) and fit the aggregation rule; teachers score the unlabeled corpora, the rule turns scores into targets $\hat y$, and one student trained on them with MSE is tested against human MOS. ASP: attentive statistics pooling.}
\label{fig:pipeline}
\end{figure}

\subsection{Overview}
Fig.~\ref{fig:pipeline} summarises the four stages: (i) benchmark $M$ public predictors on labeled data and keep the $K$ most informative; (ii) fit an aggregation rule $g_\theta$ that maps the $K$ teacher scores of an utterance to one target; (iii) apply the rule to large unlabeled corpora; (iv) train a student on the resulting targets with a mean-squared-error (MSE) loss.
Only stages (i) and (ii) touch human ratings, and only through the small public sets.

\subsection{Teacher pool and selection}
\label{sec:pool}
We consider eight open non-intrusive predictors that differ in features and training labels: MOSNet~\cite{mosnet} (spectrogram CNN-BLSTM), NISQA~\cite{nisqa}, DNSMOS~\cite{dnsmos} (P.835), HuBERT-MOS~\cite{sslmos}, XLS-R-SQA~\cite{xlsrsqa}, a regressor on XLS-R~\cite{xlsr} features, UTMOSv2~\cite{utmosv2}, Distill-MOS~\cite{distillmos} and WhiSQA~\cite{whisqa}.
Each is run with its public checkpoint and default preprocessing.
Because their output scales differ, predictions are z-scored to the mean and standard deviation of the human MOS of the development set before RMSE and MAE are computed; this affine map leaves rank and linear correlation unchanged.
Teachers are ranked by Spearman $\rho$ on URGENT and the top five (WhiSQA, Distill-MOS, UTMOSv2, XLS-R and MOSNet, abbreviated W, D, U, X, M) enter the ensemble search.
MOSNet is kept deliberately as a weak member to test whether each aggregation rule can ignore it.

\subsection{Aggregation rules}
\label{sec:agg}
For an utterance with teacher scores $\mathbf{y}=(y_1,\dots,y_K)$ the target is $\hat y = \sum_{i} w_i\,y_i$ with $\sum_i w_i=1$.
We compare five ways of setting $\mathbf{w}$:
\textbf{uniform} $w_i=1/K$;
\textbf{$\rho$-weighted} $w_i\propto\rho_i$, the teacher's Spearman correlation on the development set;
\textbf{1/RMSE-weighted} $w_i\propto 1/\mathrm{RMSE}_i$;
\textbf{MSE-optimised}, fixed weights that minimise squared error against human MOS on the labeled sets, an upper bound for any static rule;
and an \textbf{adaptive router} $\mathbf{w}=g_\theta(\mathbf{y})$, a two-layer MLP with normalised output that reads the $K$ scores and emits per-utterance weights, trained with MSE on SOMOS and BVCC only.
The router can learn which teacher to trust for which kind of input, e.g.\ a TTS-specialised model for clean synthetic speech and an enhancement-specialised model for noisy audio, without any acoustic front-end of its own.
Its input is the $K$ scores alone, so it adds negligible cost.
All $2^5-6=26$ subsets of the five teachers with $K\ge2$ are evaluated under every rule.

\subsection{Student: \proxy{}}
\label{sec:arch}
The student must be cheap at inference and multilingual, since the targets span dozens of languages.
We use the OmniASR-W2V-300M wav2vec~2.0 encoder~\cite{omniasr}, pre-trained on speech from over 1{,}600 languages, followed by attentive statistics pooling~\cite{attentive_pooling}, which attends over time and concatenates the weighted mean and standard deviation of the frame features, and a two-layer head (hidden size 1{,}024, GELU) that outputs one scalar.
The encoder is fine-tuned with a learning rate $100\times$ smaller than the head.
Training uses AdamW ($10^{-4}$ for pooling and head, $10^{-6}$ for the encoder), batch size 8, five epochs with two epochs of linear warm-up followed by cosine decay, gradient clipping at 1.0 and bfloat16 mixed precision on two RTX~4070~Ti GPUs.
The loss is MSE between the prediction and the ensemble target.
After training the model is exported to ONNX in FP32 and FP16 for deployment; no teacher is needed at test time.

\section{Data}
\label{sec:data}
\begin{table}[t]
\centering\small
\setlength{\tabcolsep}{3pt}
\caption{Corpora. Human MOS is used only for teacher ranking, router fitting and evaluation; unlabeled corpora receive ensemble targets.}
\label{tab:data}
\begin{tabular}{@{}llrl@{}}
\toprule
Corpus & Content & Utt. & Role \\
\midrule
URGENT~\cite{urgent2024}   & enhanced natural, En & 6{,}900   & rank, test \\
BVCC~\cite{bvcc}           & TTS / VC, En          & 3{,}000   & router \\
SOMOS~\cite{somos}         & neural TTS, En        & 20{,}000  & router \\
\mosds{} (ours)            & TTS / codecs, Ru      & 4{,}600   & held-out test \\
\midrule
MLAAD~\cite{mlaad}         & TTS / VC, multilingual & 200{,}000 & targets \\
Balalaika~\cite{borodin2026balalaikadatacentricprosodyawareannotation} & natural speech, Ru & 440{,}000 & targets \\
SpeechFake~\cite{speechfake} & real + synthetic    & 167{,}573 & targets \\
\bottomrule
\end{tabular}
\end{table}

Table~\ref{tab:data} lists the corpora.
URGENT~\cite{urgent2024} provides MOS for enhanced speech under additive noise, reverberation and band limitation and serves as the development set for teacher ranking; BVCC~\cite{bvcc} and SOMOS~\cite{somos} add synthetic-speech ratings for router training.
The three unlabeled corpora contribute 807k utterances: synthetic speech from many generators and languages (MLAAD, SpeechFake), the regime in which the teachers disagree most and ensemble targets are most valuable, and natural Russian speech from the OpenSTT-derived subset of the Balalaika corpus~\cite{borodin2026balalaikadatacentricprosodyawareannotation}, the language of the held-out benchmark.
\mosds{} is never used for selection, fitting or training.

\subsection{The \mosds{} benchmark}
\label{sec:mos260}
\mosds{} contains 4{,}600 Russian utterances from 38 synthesis conditions: a natural reference; four neural codecs (EnCodec~\cite{encodec}, SpeechTokenizer~\cite{speechtokenizer}, SNAC~\cite{snac}, WavTokenizer~\cite{wavtokenizer}); six vocoder front-ends pairing mel or deep-compression autoencoder (DC-AE~\cite{chen2025dc}) features with frozen BigVGAN, fine-tuned BigVGAN~\cite{bigvgan} or Vocos~\cite{vocos}; five public zero-shot systems (F5-TTS~\cite{f5tts}, XTTS, FishSpeech, Tortoise and GPT-SoVITS); six Grad-TTS~\cite{gradtts} encoder-curriculum variants from~\cite{borodin2026prosody}; and 16 configurations of an in-house multi-stage diffusion TTS system~\cite{borodin2026prosody}.
Ratings were collected on the open-source LabelSpeech platform~\cite{labelspeech} from volunteer native Russian speakers who gave informed consent, on a 0--5 quality scale (0 non-speech, 5 studio quality) and a parallel 0--5 intonation scale; every utterance was rated by at least seven listeners before filtering.
Raters were filtered post hoc: leave-one-out Pearson correlation with the other raters below $0.3$, per-item rating variance in the bottom or top 5\% of raters, or a bias $z$-score $\ge2.5$ removed a rater and all of their ratings.
The final scores are means over the remaining raters: 11{,}545 quality ratings on 4{,}355 utterances (median 3 per utterance) and 28{,}068 intonation ratings on 4{,}598 utterances (median 6); the other 245 utterances were rated for intonation only. Quality MOS averages $3.34\pm1.14$ and correlates only $\rho=0.52$ with intonation MOS. The 4{,}355 quality-rated utterances form the evaluation set.

\section{Experiments}
\label{sec:exp}
\subsection{Metrics and uncertainty}
We report Pearson $r$, Spearman $\rho$, Kendall $\tau$, RMSE and MAE at the utterance level, with $\rho$ as the primary metric because MOS is ordinal.
Confidence intervals for $\rho$ use the Fisher $z$ approximation with the Spearman correction of~\cite{bonett2000}; with $N{=}6{,}900$ (URGENT) and $N{=}4{,}355$ (\mosds{}) the 95\% half-width is at most $0.010$ and $0.025$, respectively, for $\rho\ge0.45$.

\subsection{Teachers across domains}
\label{sec:teachers}
\begin{table*}[t]
\centering\small
\setlength{\tabcolsep}{5pt}
\caption{Utterance-level agreement with human MOS. Predictions are z-score aligned before RMSE/MAE. Best value per column in bold. Ensemble rows are the four-model teacher ensemble used to label the training corpora; the router is trained on SOMOS and BVCC, so both benchmarks are held out for every row. 95\% CI half-width for $\rho$: $\le0.010$ (URGENT), $\le0.025$ (\mosds{}).}
\label{tab:main}
\begin{tabular}{@{}l ccccc ccccc@{}}
\toprule
& \multicolumn{5}{c}{URGENT ($N{=}6{,}900$)} & \multicolumn{5}{c}{\mosds{} ($N{=}4{,}355$)} \\
\cmidrule(lr){2-6}\cmidrule(lr){7-11}
Model & $r\uparrow$ & $\rho\uparrow$ & $\tau\uparrow$ & RMSE$\downarrow$ & MAE$\downarrow$ & $r\uparrow$ & $\rho\uparrow$ & $\tau\uparrow$ & RMSE$\downarrow$ & MAE$\downarrow$ \\
\midrule
\proxy{} (PyTorch)        & \best{0.806} & \best{0.802} & \best{0.620} & \best{0.471} & \best{0.368} & 0.691 & 0.636 & 0.474 & 0.897 & 0.692 \\
\proxy{} (ONNX FP32)      & 0.779 & 0.778 & 0.591 & 0.502 & 0.393 & \best{0.700} & \best{0.647} & \best{0.481} & \best{0.883} & \best{0.687} \\
\proxy{} (ONNX FP16)      & 0.779 & 0.778 & 0.591 & 0.502 & 0.393 & \best{0.700} & \best{0.647} & \best{0.481} & \best{0.883} & \best{0.687} \\
\midrule
Ensemble W+D+U+X, uniform & 0.813 & 0.814 & 0.629 & 0.448 & 0.356 & 0.674 & 0.578 & 0.420 & 0.886 & 0.701 \\
Ensemble W+D+U+X, router  & 0.819 & 0.816 & 0.632 & 0.439 & 0.348 & 0.706 & 0.636 & 0.471 & 0.822 & 0.633 \\
\midrule
WhiSQA (W)~\cite{whisqa}       & 0.780 & 0.773 & 0.588 & 0.502 & 0.397 & 0.576 & 0.466 & 0.332 & 1.050 & 0.841 \\
Distill-MOS (D)~\cite{distillmos} & 0.759 & 0.748 & 0.568 & 0.525 & 0.411 & 0.670 & 0.613 & 0.456 & 0.927 & 0.712 \\
UTMOSv2 (U)~\cite{utmosv2}     & 0.721 & 0.748 & 0.560 & 0.565 & 0.450 & 0.622 & 0.555 & 0.403 & 0.991 & 0.781 \\
XLS-R-SQA (X)~\cite{xlsrsqa} & 0.743 & 0.745 & 0.561 & 0.542 & 0.426 & 0.583 & 0.473 & 0.337 & 1.041 & 0.834 \\
MOSNet (M)~\cite{mosnet}       & 0.279 & 0.110 & 0.073 & 0.908 & 0.702 & $-$0.210 & $-$0.209 & $-$0.146 & 1.773 & 1.384 \\
NISQA~\cite{nisqa}             & 0.028 & $-$0.027 & $-$0.021 & 1.055 & 0.863 & 0.261 & 0.226 & 0.158 & 1.386 & 1.105 \\
DNSMOS~\cite{dnsmos}           & $-$0.010 & $-$0.041 & $-$0.028 & 1.075 & 0.876 & 0.075 & 0.056 & 0.039 & 1.551 & 1.254 \\
HuBERT-MOS~\cite{sslmos}       & $-$0.005 & $-$0.002 & $-$0.001 & 1.073 & 0.855 & $-$0.003 & $-$0.009 & $-$0.006 & 1.614 & 1.294 \\
\bottomrule
\end{tabular}
\end{table*}

The lower block of Table~\ref{tab:main} benchmarks the teachers.
Four SSL-based models form a tight leading group on URGENT ($\rho$ from $0.745$ to $0.773$), while MOSNet, NISQA, DNSMOS and HuBERT-MOS are near or below zero: models trained for telephony or enhancement do not transfer to the URGENT mixture of distortions, and the plain HuBERT regressor collapses entirely.
On \mosds{} the ranking flips.
Distill-MOS leads ($\rho=0.613$), UTMOSv2 follows, and WhiSQA, the best model on URGENT, drops to $0.466$; MOSNet becomes anti-correlated with human judgment.
The WhiSQA--Distill-MOS gap is $0.025$ on URGENT and $0.147$ on \mosds{}, more than twice the CI half-width in each case, so the flip is not noise.
No single public predictor is reliable in both domains, which is the premise of the ensemble.

\subsection{Ensemble search}
\label{sec:ensembles}
\begin{table}[t]
\centering\small
\setlength{\tabcolsep}{2.2pt}
\caption{Best Spearman $\rho$ over all teacher subsets of size $K$ under each aggregation rule. MSE-optimised weights are fit on URGENT (in-sample there); the router is trained on SOMOS and BVCC, so both benchmarks are held out for it. Winning subsets at $K{=}4$: W+D+U+X for every rule except $\rho$-weighting on \mosds{} (D+U+X+M); D+U ($K{=}2$) and D+U+X ($K{=}3$) on \mosds{}.}
\label{tab:ens}
\begin{tabular}{@{}l cccc cccc@{}}
\toprule
& \multicolumn{4}{c}{URGENT} & \multicolumn{4}{c}{\mosds{}} \\
\cmidrule(lr){2-5}\cmidrule(lr){6-9}
Rule & $K{=}2$ & 3 & 4 & 5 & $K{=}2$ & 3 & 4 & 5 \\
\midrule
Best single & \multicolumn{4}{c}{0.773 (W)} & \multicolumn{4}{c}{0.613 (D)} \\
Uniform mean      & .799 & .809 & .814 & .804 & .617 & .602 & .578 & .510 \\
$\rho$-weighted   & .799 & .809 & .814 & .814 & .619 & .608 & .608 & .588 \\
1/RMSE-weighted   & .799 & .809 & .814 & .811 & .619 & .605 & .583 & .547 \\
MSE-optimised     & .795 & .809 & .813 & .811 & .617 & .602 & .578 & .510 \\
Adaptive router   & .795 & \best{.810} & \best{.816} & \best{.817} & \best{.635} & \best{.635} & \best{.636} & \best{.636} \\
\bottomrule
\end{tabular}
\end{table}

\begin{figure}[t]
\centering
\includegraphics[width=\columnwidth]{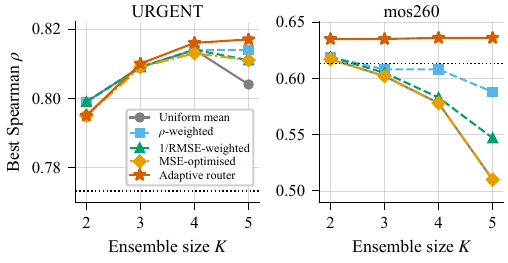}
\vspace{-8pt}
\caption{Best $\rho$ versus ensemble size for each aggregation rule (Table~\ref{tab:ens}); dotted line: best single teacher. Static rules peak at $K{=}4$ on URGENT and decay on \mosds{} as weaker teachers enter; the router is flat or rising in both domains.}
\label{fig:ens}
\end{figure}

Table~\ref{tab:ens} and Fig.~\ref{fig:ens} summarise the 26-subset search.
Three findings stand out.
(1) \textit{Averaging beats every teacher.} On URGENT the uniform mean of WhiSQA, Distill-MOS, UTMOSv2 and XLS-R reaches $\rho=0.814$, $+0.041$ over WhiSQA, and every one of the ten best subsets contains at least two of W, U and X, so the gain comes from complementary errors rather than from any one model.
(2) \textit{Static weights barely help.} $\rho$-weighted, 1/RMSE-weighted and MSE-optimised rules land within $0.003$ of the uniform mean at $K{=}4$ on URGENT and do not rescue \mosds{}, where the best static subset is the pair Distill-MOS + UTMOSv2 ($0.617$) and adding WhiSQA or MOSNet lowers $\rho$ monotonically to $0.510$ at $K{=}5$: a domain-mismatched member cannot be down-weighted by a constant.
(3) \textit{Routing is robust.} The per-utterance router holds $\rho\approx0.636$ on \mosds{} for every $K$ from 2 to 5, i.e.\ it learns to switch off WhiSQA and MOSNet on synthetic Russian speech although it saw neither benchmark during training, and it is the only rule whose URGENT score still rises at $K{=}5$ ($0.817$).
The full search makes the same point at the subset level.
Across the ten available pairs of subsets that differ only by MOSNet, adding it lowers $\rho$ on average by $0.024$ on URGENT and $0.115$ on \mosds{} under the uniform mean, but by only $0.001$ and $0.007$ under the router, and the routed five-model set ranks first on both benchmarks.
Under the router the best two-teacher subset on \mosds{} (D+U, $0.635$) already matches the five-teacher one, so a two-teacher router is a viable low-cost target generator.
For the W+D+U+X subset the four static rules are within $0.011$ of one another on every metric, whereas the router lowers RMSE by $7\%$ and raises $r$ by $0.03$ on \mosds{} (Table~\ref{tab:main}).
We therefore use the routed W+D+U+X ensemble to label the unlabeled corpora; the fifth member adds cost without benefit on the held-out domain.

\subsection{Distilled student}
\label{sec:student}
The upper block of Table~\ref{tab:main} evaluates \proxy{}.
On URGENT the student reaches $\rho=0.802$ (95\% CI $[0.793,0.811]$), above every teacher (WhiSQA $[0.763,0.783]$) with $6\%$ lower RMSE; on \mosds{} it reaches $0.636$ ($[0.616,0.654]$) against $0.613$ for Distill-MOS, a smaller margin whose interval overlaps the teacher's.
Relative to the routed teacher ensemble it was trained to imitate, the student keeps $67\%$ of the ensemble's gain over the best teacher on URGENT and $100\%$ on \mosds{}, while running one 300M-parameter encoder instead of four teachers plus a router.
The ONNX exports change $\rho$ by at most $0.024$ and FP16 is lossless relative to FP32; on \mosds{} the export is even slightly better ($0.647$), so the deployable graph is the strongest single model on synthetic speech.
Two effects plausibly combine here: the ensemble target is smoother than any single teacher, and the student sees 807k diverse utterances, two orders of magnitude more than the human-rated sets its teachers were trained on, 440k of them in Russian, the language of \mosds{}.
We do not claim to separate these effects.

\subsection{System-level behaviour on \mosds{}}
\label{sec:systems}
\begin{figure}[t]
\centering
\includegraphics[width=\columnwidth]{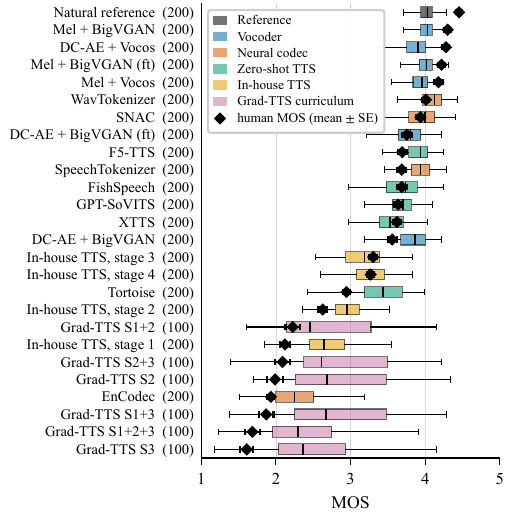}
\vspace{-8pt}
\caption{\mosds{} per synthesis condition, sorted by human MOS: boxes show the distribution of \proxy{} scores (quartiles, $1.5\,$IQR whiskers; counts in brackets; the 16 in-house configurations pooled by stage), diamonds the human MOS mean $\pm$ SE.}
\label{fig:systems}
\end{figure}
Fig.~\ref{fig:systems} overlays human MOS on the \proxy{} score distributions per condition; over the 38 conditions the means agree at $\rho=0.951$ and $r=0.948$ ($0.944$ and $0.962$ over the 26 shown).
The student reproduces the human ordering of the six families exactly (Table~\ref{tab:families}) and the design intent of each condition: mel-vocoder resyntheses sit just below the natural reference, EnCodec is the worst codec, and the in-house TTS improves from stage 1 to stage 3 in both scores.
Its errors are compressions at the extremes: it over-scores Tortoise ($3.43$ vs $2.94$) and the Grad-TTS variants ($2.4$--$2.9$ vs $1.6$--$2.2$), and it cannot separate the five best conditions, which humans spread over $4.0$--$4.45$; predictions are compressed towards the middle of the scale (the reference scores $3.99$ against a human $4.45$, the worst Grad-TTS variant $2.48$ against $1.61$), so absolute calibration rather than ranking is the student's weak point.
\proxy{} correlates $\rho=0.50$ with the intonation scores, close to the $0.52$ between the two human scores, so it tracks acoustic quality rather than prosody.
Per-condition human and predicted scores are released with the benchmark.

\begin{table}[t]
\centering\small
\setlength{\tabcolsep}{2.5pt}
\caption{\mosds{} by family of synthesis condition: human MOS and \proxy{} score (mean $\pm$ SD over utterances).}
\label{tab:families}
\begin{tabular}{@{}l rrcc@{}}
\toprule
Family & Cond. & Utt. & Human MOS & \proxy{} \\
\midrule
Natural reference    & 1  & 200   & $4.45\pm0.62$ & $3.99\pm0.17$ \\
Vocoder front-ends   & 6  & 1{,}200 & $4.05\pm0.78$ & $3.88\pm0.24$ \\
Zero-shot TTS        & 5  & 1{,}000 & $3.52\pm0.80$ & $3.63\pm0.33$ \\
Neural codecs        & 4  & 800   & $3.39\pm1.11$ & $3.54\pm0.78$ \\
In-house TTS (4 stages) & 16 & 800 & $2.83\pm0.90$ & $3.02\pm0.37$ \\
Grad-TTS curriculum  & 6  & 600   & $1.91\pm0.98$ & $2.69\pm0.77$ \\
\bottomrule
\end{tabular}
\end{table}

\subsection{Limitations}
The static rules use URGENT statistics, so their URGENT entries in Table~\ref{tab:ens} are in-sample; the router, every \mosds{} number and the student are held out.
A median of only three quality ratings per utterance survives filtering, which caps utterance-level agreement on \mosds{}, whereas condition means pool at least 94 utterances; the released benchmark carries the rater count of every utterance so that users can weight items accordingly.
\mosds{} covers one language and the pool one predictor per family.
Targets inherit the ensemble's ceiling; confidence-weighted targets based on teacher disagreement are a natural next step.

\section{Conclusion}
\label{sec:conclusion}
Public MOS predictors disagree across domains, but their disagreement is exploitable.
A learned per-utterance router over four of them is robust to weak members, and a single wav2vec~2.0 student distilled from 807k routed targets outperforms every teacher on URGENT and on the new Russian TTS benchmark \mosds{}, at the cost of one forward pass.
Ensemble supervision on unlabeled speech is thus a practical route to scalable quality assessment.
Three recommendations follow from the search: combine at least three SSL-based predictors with complementary training labels, since every strong subset paired Distill-MOS with two of WhiSQA, UTMOSv2 and XLS-R-SQA; prefer a learned router over static weights when the deployment domain is unknown, because only the router survived a weak member; and distill the ensemble rather than deploying it, since the student kept its held-out accuracy; model, code, ONNX exports and \mosds{} are publicly released.\footnotemark[1]
Future work will weight targets by teacher disagreement, add per-rater releases and further languages to \mosds{}, and report system-level agreement on URGENT.

\needspace{5\baselineskip}\vspace{8pt}\centerline{\bfseries ACKNOWLEDGEMENTS}\vspace{5pt}
\noindent We used Claude Fable 5 (Anthropic) for text polishing only. All ideas, methods and experiments are the authors' own.

\needspace{8\baselineskip}\vspace{8pt}\centerline{\bfseries COMPLIANCE WITH ETHICAL STANDARDS}\vspace{5pt}
\noindent\mosds{} ratings were collected through the open-source LabelSpeech platform~\cite{labelspeech}; all raters were volunteers who gave informed consent, and the release contains audio and aggregated scores only. All other corpora and models are public research resources. The authors declare no competing interests and received no funding.

\balance
\bibliographystyle{IEEEbib}
\bibliography{references}

\clearpage
\nobalance
\onecolumn
\twocolumn[{\centering\Large\bfseries SUPPLEMENTARY MATERIAL\par\bigskip}]
\appendix
\setcounter{table}{0}\setcounter{figure}{0}
\renewcommand{\thetable}{S\arabic{table}}\renewcommand{\theHtable}{S\arabic{table}}
\renewcommand{\thefigure}{S\arabic{figure}}\renewcommand{\theHfigure}{S\arabic{figure}}

\setcounter{topnumber}{4}\setcounter{dbltopnumber}{4}\setcounter{totalnumber}{8}
\renewcommand{\topfraction}{.95}\renewcommand{\dbltopfraction}{.85}
\renewcommand{\textfraction}{.05}
\renewcommand{\floatpagefraction}{.75}\renewcommand{\dblfloatpagefraction}{.75}
\makeatletter
\setlength{\@fptop}{0pt}\setlength{\@fpbot}{0pt plus 1fil}
\setlength{\@dblfptop}{0pt}\setlength{\@dblfpbot}{0pt plus 1fil}
\newcommand{\inlinecaption}[1]{\refstepcounter{table}\@makecaption{\fnum@table}{#1}\vskip\belowcaptionskip}
\makeatother

\section{Contents}
This supplement expands the main paper with (i) additional figures (Figs.~\ref{fig:supp_teachers}--\ref{fig:supp_top_mos260}), (ii) the full \mosds{} rating rubrics and annotator-filtering procedure, (iii) the \proxy{} training configuration, (iv) confidence intervals for every model (Table~\ref{tab:supp_ci}), and (v) every teacher subset under every aggregation rule on both benchmarks (Tables~\ref{tab:supp_mean}--\ref{tab:supp_router}).

\section{Additional figures}
\begin{figure}[t]\centering\includegraphics[width=\columnwidth]{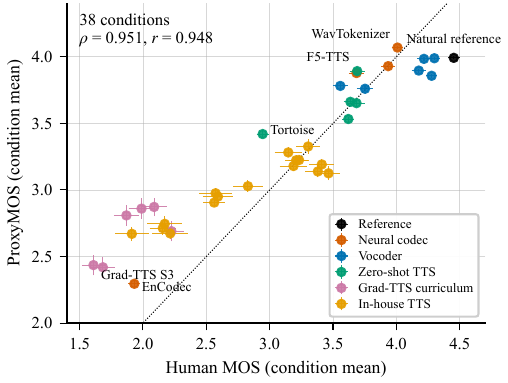}
\caption{Condition-level agreement on \mosds{}: mean human MOS versus mean \proxy{} score for the 38 synthesis conditions (bars: standard error; dotted line: identity). Table~\ref{tab:supp_conditions} gives the scores per condition.}\label{fig:supp_scatter}\end{figure}
\begin{figure}[t]\centering\includegraphics[width=\columnwidth]{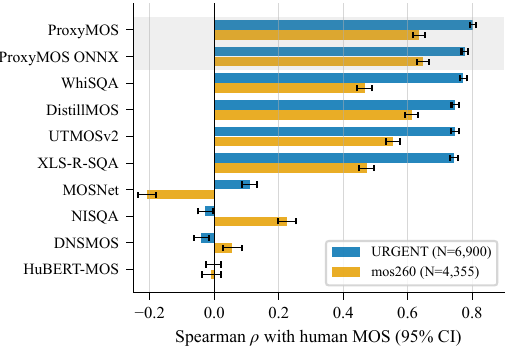}
\caption{Spearman $\rho$ of every model with 95\% confidence intervals on URGENT and \mosds{}. Shaded rows: the distilled student.}\label{fig:supp_teachers}\end{figure}
\begin{figure}[t]\centering\includegraphics[width=\columnwidth]{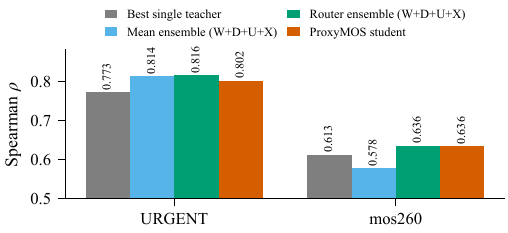}
\caption{Best single teacher, the W+D+U+X ensemble under uniform and routed aggregation, and the distilled student (PyTorch), Spearman $\rho$ on both benchmarks.}\label{fig:supp_gain}\end{figure}
\begin{figure}[t]\centering\includegraphics[width=\columnwidth]{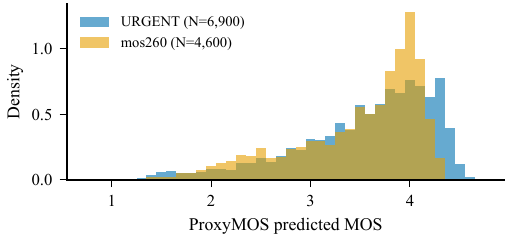}
\caption{Distribution of \proxy{} predictions on the two benchmarks.}\label{fig:supp_hist}\end{figure}
\begin{figure*}[t]\centering\includegraphics[width=\textwidth]{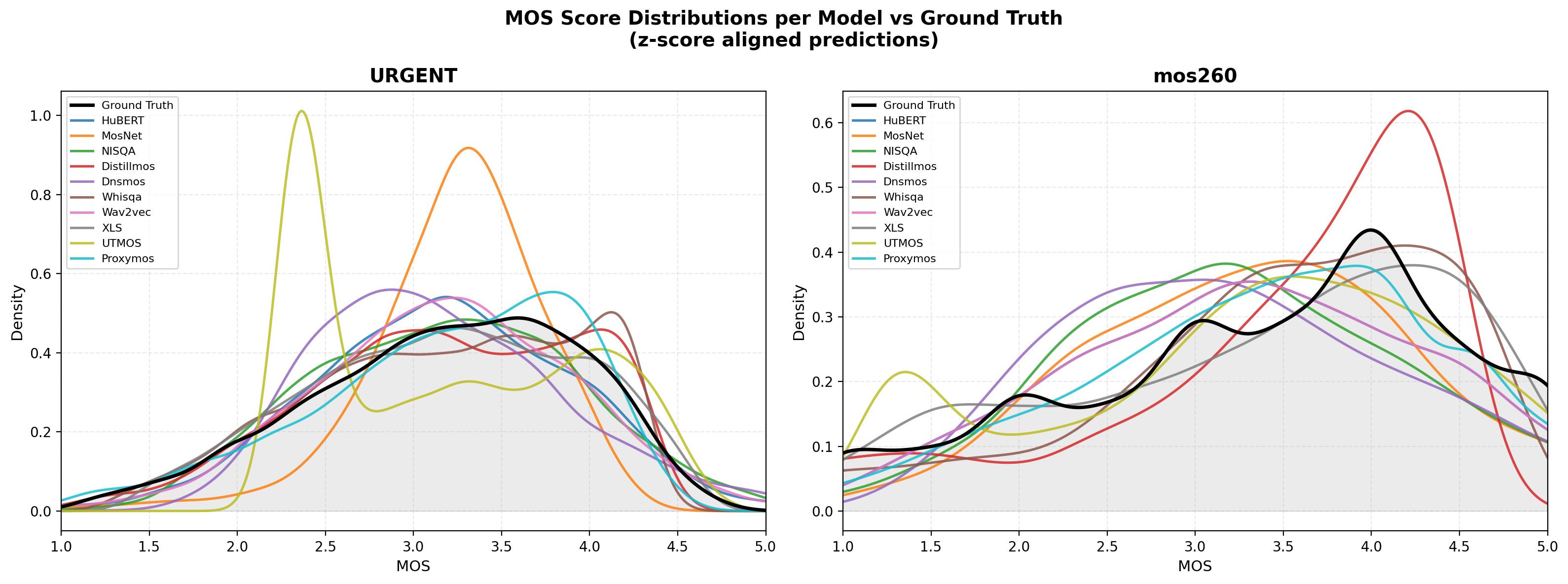}
\caption{Kernel density of z-score-aligned predictions of every model against the human MOS distribution (black) on URGENT (left) and \mosds{} (right).}\label{fig:supp_kde}\end{figure*}
\begin{figure*}[t]\centering\includegraphics[width=\textwidth]{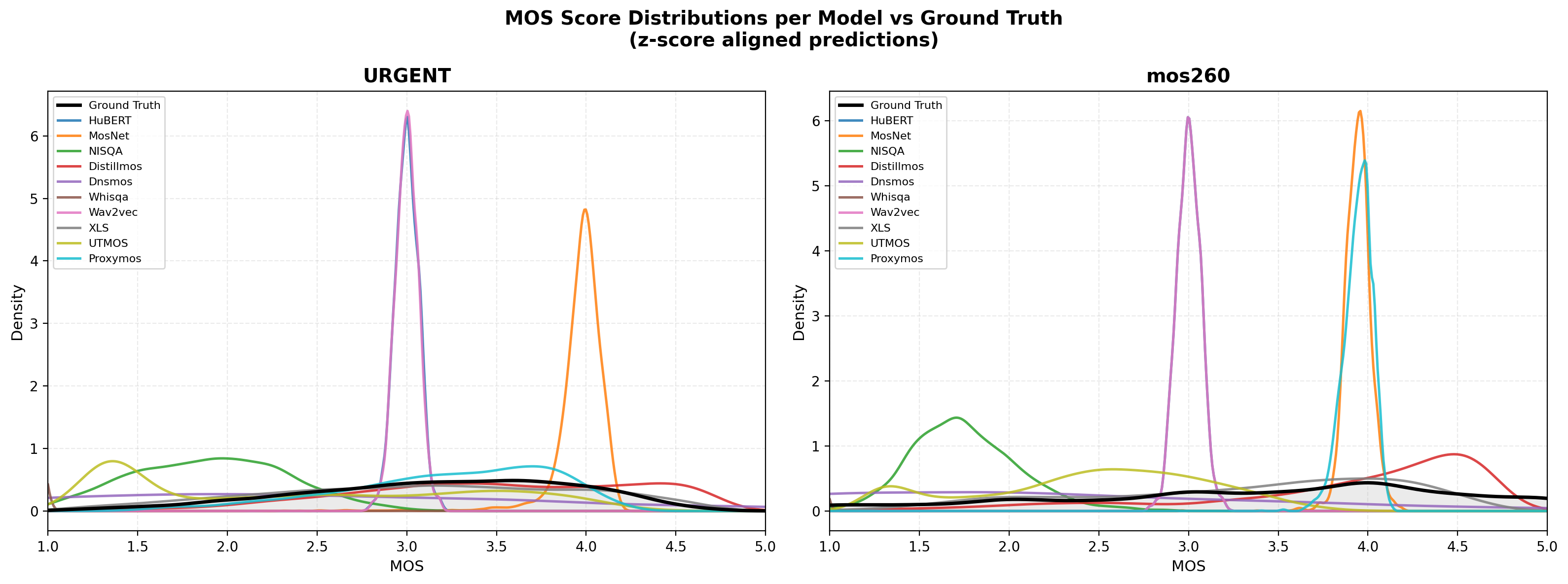}
\caption{Kernel density of raw (unaligned) predictions of every model on URGENT and \mosds{}. Several models (HuBERT-MOS, DNSMOS, MOSNet) collapse to a narrow band, which explains their near-zero correlation.}\label{fig:supp_kde_raw}\end{figure*}
\begin{figure*}[t]\centering\includegraphics[width=\textwidth]{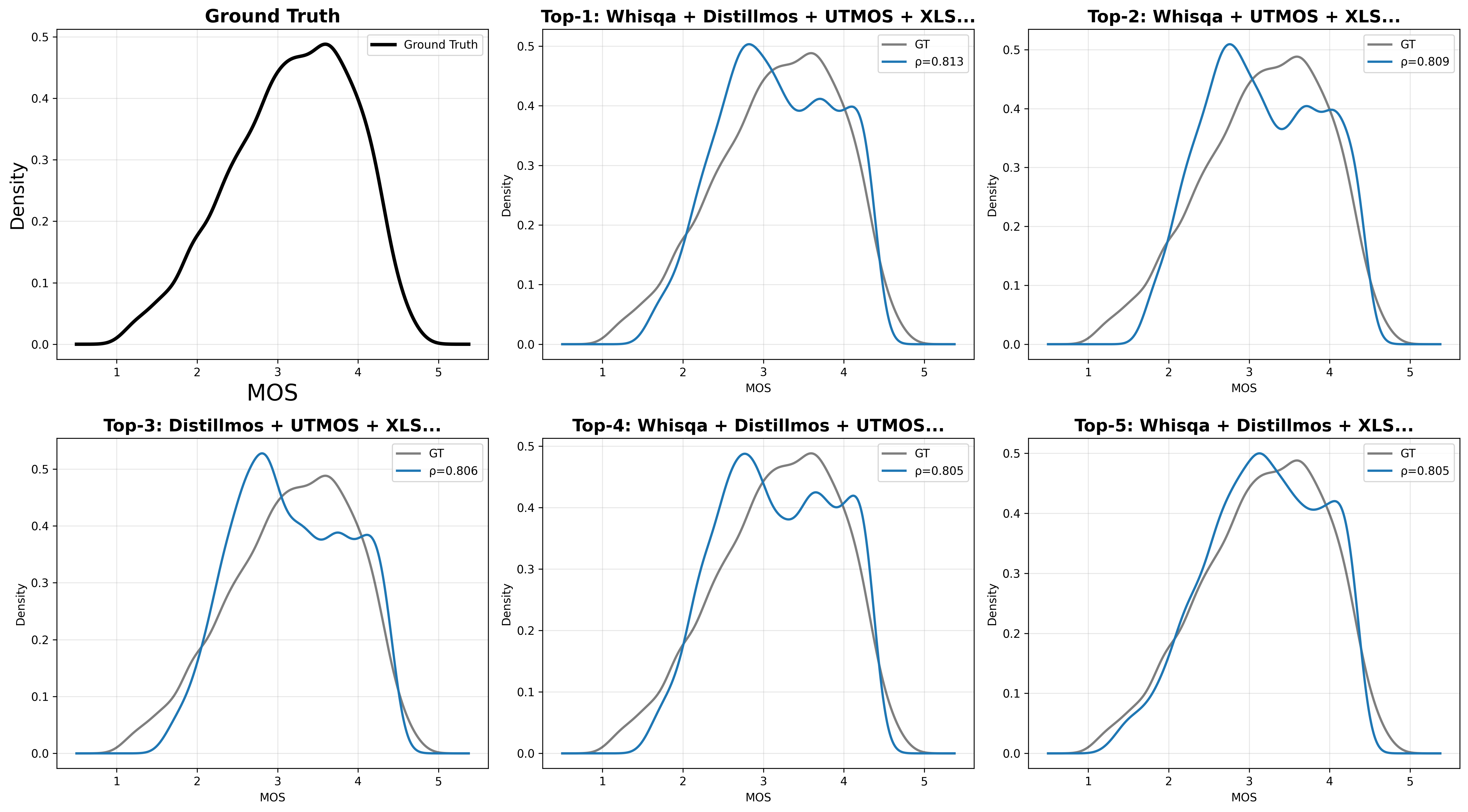}
\caption{URGENT: score distributions of the five best uniform-mean ensembles after z-score alignment (blue) against human MOS (grey).}\label{fig:supp_top_urgent}\end{figure*}
\begin{figure*}[t]\centering\includegraphics[width=\textwidth]{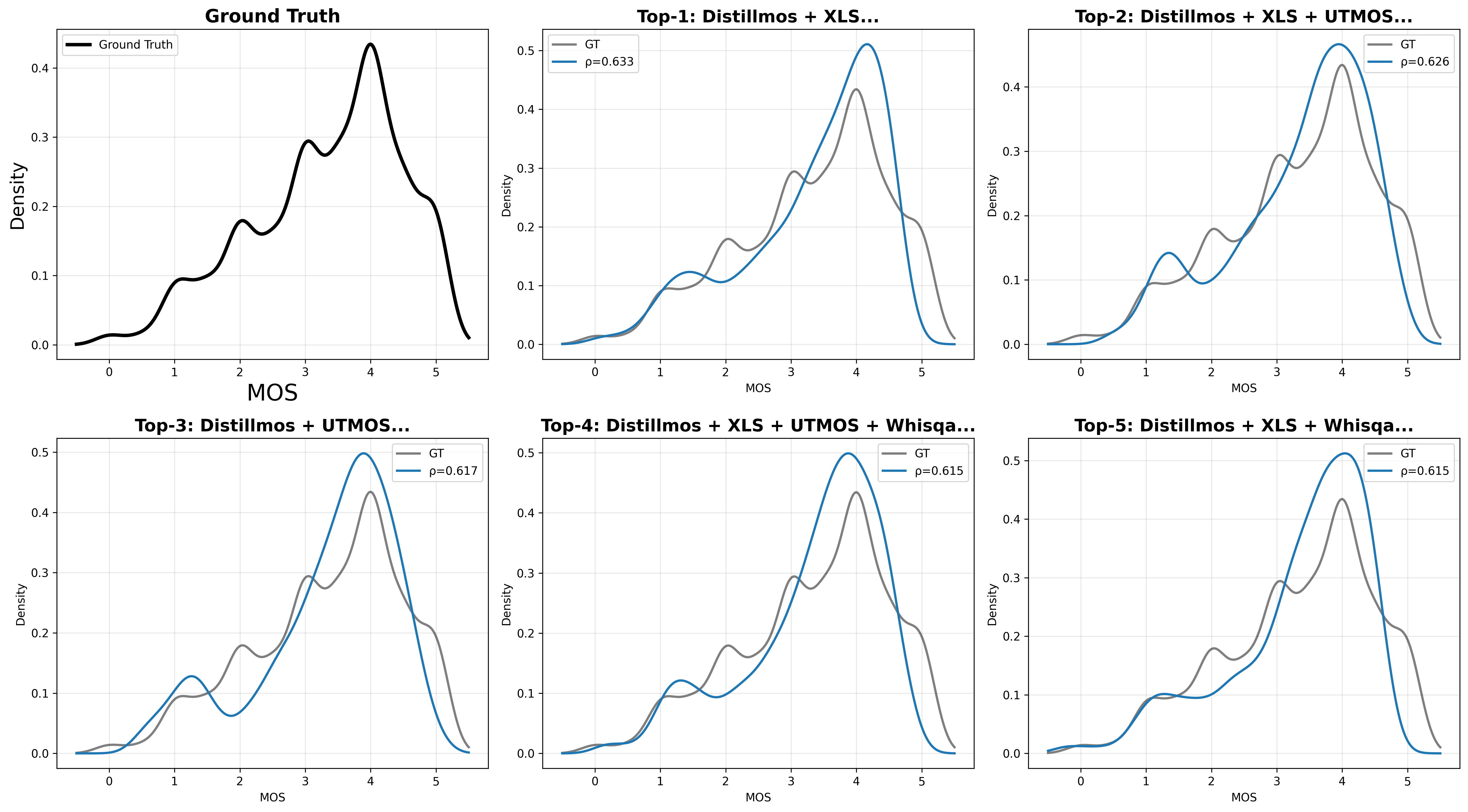}
\caption{\mosds{}: score distributions of the five best uniform-mean ensembles after z-score alignment against human MOS.}\label{fig:supp_top_mos260}\end{figure*}

Fig.~\ref{fig:supp_teachers} adds confidence intervals to the teacher benchmark, Fig.~\ref{fig:supp_gain} summarises teacher, ensemble and student accuracy, Fig.~\ref{fig:supp_hist} shows the distribution of student predictions, Figs.~\ref{fig:supp_kde} and~\ref{fig:supp_kde_raw} compare aligned and raw prediction densities of every model with the human MOS distribution, and Figs.~\ref{fig:supp_top_urgent} and~\ref{fig:supp_top_mos260} show the five best uniform-mean ensembles per benchmark.

\section{\mosds{} per condition}
\input{supp/mos260_conditions.tex}
Table~\ref{tab:supp_conditions} lists, for every condition, the human quality and intonation scores and the \proxy{} score; Fig.~\ref{fig:supp_scatter} plots condition means against each other.

\section{\mosds{} rating protocol}
\noindent\textbf{Quality (MOS) rubric.}
5: perfect studio quality, no noise, reverberation, distortion or robotic voice.
4: studio quality with minor artefacts (slight noise or reverberation), speech clear.
3: unprofessional recording: noticeable noise, distortion, poor clarity or background music.
2: very low quality, severe distortion or noise typical of telephony.
1: low-quality telephony, speech barely understandable, intermittent sound.
0: non-speech (white noise, silence, unrecognisable sounds).
\smallskip\noindent\textbf{Intonation (IntMOS) rubric.}
5: sounds like a person in normal conversation, correct stresses and logical pauses.
4: dictation- or audiobook-like delivery, recognisably human, no stress errors.
3: indeterminate, individual stress or pause errors, broken intonation structure.
2: mostly robotic, frequent stress and pause errors.
1: clearly artificial, most stresses and pauses wrong.
0: unintelligible, intonation cannot be judged.
\smallskip\noindent\textbf{Annotator filtering.}
All raters were native Russian speakers recruited through LabelSpeech. For each rater we computed (a) the Pearson correlation between their scores and the leave-one-out mean of all other raters over shared items (exclusion if $r<0.3$ with at least ten shared items; fewer than ten shared items triggered a warning only), (b) their per-item rating variance (exclusion if at or below the 5th percentile, ``clickers'', or at or above the 95th percentile, ``chaotic''), and (c) a bias $z$-score of their mean rating against the grand mean (exclusion if $|z|\ge2.5$). All ratings of an excluded rater were discarded and MOS was recomputed as the mean over the remaining raters. Filtering was applied to the MOS projects only; IntMOS ratings were left untouched.

\section{\proxy{} training configuration}
Table~\ref{tab:supp_hp} lists the student's architecture, optimisation and data; the targets are the routed W+D+U+X ensemble scores of the main paper.
\par\medskip\noindent\begin{minipage}{\columnwidth}\centering\small\setlength{\tabcolsep}{3pt}
\inlinecaption{Student configuration.}\label{tab:supp_hp}
\begin{tabular}{@{}ll@{}}\toprule
Encoder & OmniASR-W2V-300M (w2v2), fine-tuned \\
Pooling & attentive statistics (mean $\|$ std over time) \\
Head & Linear--GELU--Linear (1{,}024), scalar out \\
Loss & MSE to routed W+D+U+X ensemble target \\
Optimiser & AdamW; lr $10^{-4}$ (pool/head), $10^{-6}$ (encoder) \\
Schedule & 2 ep.\ linear warm-up, cosine; 5 epochs \\
Batch / precision & 8 / bfloat16 (Hugging Face Accelerate) \\
Gradient clipping & max norm 1.0 \\
Data & MLAAD, Balalaika, SpeechFake (807k utt.) \\
Hardware & 2$\times$ NVIDIA RTX 4070 Ti \\
Export & ONNX FP32 and FP16 \\
\bottomrule\end{tabular}\end{minipage}\par\medskip

\input{supp/tables_generated.tex}
\section{Confidence intervals and all ensemble subsets}
Table~\ref{tab:supp_ci} gives the Spearman correlation of every model with its 95\% confidence interval on both benchmarks.
Tables~\ref{tab:supp_mean}--\ref{tab:supp_router} list every teacher subset under each aggregation rule, URGENT and \mosds{} side by side; each side is sorted by its own Spearman $\rho$, so a row index does not denote the same subset on both sides.
Each table shows 25 of the 26 subsets: the search report lists the top 10 and the next 15.
Teachers are abbreviated W\,=\,WhiSQA, D\,=\,Distill-MOS, U\,=\,UTMOSv2, X\,=\,XLS-R, M\,=\,MOSNet; $K$ is the subset size.

\end{document}

%% file: supp/mos260_conditions.tex
\begin{table*}[t]\centering\small\setlength{\tabcolsep}{5pt}
\caption{\mosds{} per synthesis condition (in-house configurations pooled by training stage): utterances, utterances with quality ratings, human MOS, human Int-MOS and \proxy{} score (mean $\pm$ SD), sorted by human MOS.}
\label{tab:supp_conditions}
\begin{tabular}{@{}ll rr cc c@{}}\toprule Family & Condition & Utt. & Rated & Human MOS & Human Int-MOS & \proxy{} \\ \midrule
Reference & Natural reference & 200 & 190 & $4.45\pm0.62$ & $4.03\pm0.57$ & $3.99\pm0.17$ \\
Vocoder & Mel + BigVGAN & 200 & 191 & $4.30\pm0.58$ & $3.95\pm0.70$ & $3.99\pm0.18$ \\
Vocoder & DC-AE + Vocos & 200 & 190 & $4.28\pm0.56$ & $3.77\pm0.75$ & $3.86\pm0.21$ \\
Vocoder & Mel + BigVGAN (ft) & 200 & 184 & $4.22\pm0.71$ & $3.95\pm0.62$ & $3.99\pm0.17$ \\
Vocoder & Mel + Vocos & 200 & 183 & $4.18\pm0.76$ & $3.83\pm0.68$ & $3.90\pm0.20$ \\
Neural codec & WavTokenizer & 200 & 194 & $4.01\pm0.63$ & $3.54\pm0.73$ & $4.07\pm0.20$ \\
Neural codec & SNAC & 200 & 187 & $3.94\pm0.75$ & $3.59\pm0.69$ & $3.93\pm0.28$ \\
Vocoder & DC-AE + BigVGAN (ft) & 200 & 192 & $3.75\pm0.88$ & $3.67\pm0.70$ & $3.76\pm0.27$ \\
Zero-shot TTS & F5-TTS & 200 & 193 & $3.69\pm0.81$ & $3.14\pm0.71$ & $3.89\pm0.18$ \\
Neural codec & SpeechTokenizer & 200 & 185 & $3.69\pm0.81$ & $3.57\pm0.65$ & $3.88\pm0.26$ \\
Zero-shot TTS & FishSpeech & 200 & 193 & $3.69\pm0.90$ & $3.50\pm0.82$ & $3.65\pm0.38$ \\
Zero-shot TTS & GPT-SoVITS & 200 & 186 & $3.64\pm0.72$ & $3.03\pm0.68$ & $3.66\pm0.25$ \\
Zero-shot TTS & XTTS & 200 & 187 & $3.62\pm0.64$ & $3.18\pm0.67$ & $3.53\pm0.25$ \\
Vocoder & DC-AE + BigVGAN & 200 & 189 & $3.56\pm0.83$ & $3.65\pm0.71$ & $3.79\pm0.28$ \\
In-house TTS & In-house TTS, stage 3 & 200 & 195 & $3.30\pm0.69$ & $3.10\pm0.69$ & $3.17\pm0.30$ \\
In-house TTS & In-house TTS, stage 4 & 200 & 188 & $3.27\pm0.67$ & $3.14\pm0.69$ & $3.25\pm0.29$ \\
Zero-shot TTS & Tortoise & 200 & 190 & $2.94\pm0.66$ & $2.93\pm0.67$ & $3.43\pm0.34$ \\
In-house TTS & In-house TTS, stage 2 & 200 & 189 & $2.63\pm0.77$ & $2.44\pm0.79$ & $2.96\pm0.25$ \\
Grad-TTS curriculum & Grad-TTS S1+2 & 100 & 95 & $2.22\pm0.94$ & $2.45\pm0.85$ & $2.69\pm0.73$ \\
In-house TTS & In-house TTS, stage 1 & 200 & 190 & $2.12\pm0.88$ & $1.97\pm0.87$ & $2.70\pm0.35$ \\
Grad-TTS curriculum & Grad-TTS S2+3 & 100 & 95 & $2.09\pm1.00$ & $2.47\pm0.98$ & $2.88\pm0.69$ \\
Grad-TTS curriculum & Grad-TTS S2 & 100 & 95 & $1.99\pm1.00$ & $2.27\pm0.93$ & $2.87\pm0.76$ \\
Neural codec & EnCodec & 200 & 189 & $1.93\pm0.64$ & $2.54\pm0.86$ & $2.29\pm0.38$ \\
Grad-TTS curriculum & Grad-TTS S1+3 & 100 & 94 & $1.87\pm0.95$ & $2.20\pm0.85$ & $2.81\pm0.76$ \\
Grad-TTS curriculum & Grad-TTS S1+2+3 & 100 & 97 & $1.68\pm0.97$ & $2.06\pm0.98$ & $2.42\pm0.79$ \\
Grad-TTS curriculum & Grad-TTS S3 & 100 & 94 & $1.61\pm0.86$ & $2.09\pm0.94$ & $2.48\pm0.77$ \\
\bottomrule\end{tabular}\end{table*}

%% file: supp/tables_generated.tex
\begin{table*}[t]\centering\small\setlength{\tabcolsep}{6pt}
\caption{Spearman $\rho$ with 95\% confidence intervals (Fisher $z$ with the Spearman correction of Bonett and Wright) for every model on both benchmarks.}
\label{tab:supp_ci}
\begin{tabular}{@{}l cc cc@{}}\toprule & \multicolumn{2}{c}{URGENT} & \multicolumn{2}{c}{mos260} \\ \cmidrule(lr){2-3}\cmidrule(lr){4-5} Model & $\rho$ & 95\% CI & $\rho$ & 95\% CI \\ \midrule
ProxyMOS (PyTorch) & 0.802 & [0.792, 0.812] & 0.636 & [0.616, 0.655] \\
ProxyMOS (ONNX FP32) & 0.778 & [0.767, 0.788] & 0.647 & [0.628, 0.666] \\
ProxyMOS (ONNX FP16) & 0.778 & [0.767, 0.788] & 0.647 & [0.628, 0.666] \\
WhiSQA & 0.773 & [0.762, 0.784] & 0.466 & [0.442, 0.491] \\
DistillMOS & 0.748 & [0.736, 0.760] & 0.613 & [0.593, 0.633] \\
UTMOSv2 & 0.748 & [0.736, 0.759] & 0.555 & [0.532, 0.576] \\
XLS-R-SQA & 0.745 & [0.733, 0.756] & 0.473 & [0.448, 0.497] \\
MOSNet & 0.110 & [0.087, 0.133] & -0.209 & [-0.237, -0.180] \\
NISQA & -0.027 & [-0.051, -0.004] & 0.226 & [0.197, 0.254] \\
DNSMOS & -0.041 & [-0.064, -0.017] & 0.056 & [0.026, 0.086] \\
HuBERT-MOS & -0.002 & [-0.026, 0.022] & -0.009 & [-0.039, 0.020] \\
\bottomrule\end{tabular}\end{table*}

\begin{table*}[t]\centering\small\setlength{\tabcolsep}{4pt}\renewcommand{\arraystretch}{0.93}
\caption{Teacher subsets under the uniform mean rule, 25 of 26, each side sorted by its own Spearman $\rho$.}
\label{tab:supp_mean}
\begin{minipage}[t]{0.48\textwidth}\centering
\begin{tabular}{@{}l r rrrrr@{}}\toprule \multicolumn{7}{@{}c@{}}{\textbf{URGENT}} \\ \cmidrule{1-7} Subset & $K$ & $r$ & $\rho$ & $\tau$ & RMSE & MAE \\ \midrule
W+D+U+X & 4 & 0.813 & 0.814 & 0.629 & 0.448 & 0.356 \\
W+U+X & 3 & 0.804 & 0.809 & 0.623 & 0.460 & 0.366 \\
D+U+X & 3 & 0.802 & 0.806 & 0.621 & 0.461 & 0.366 \\
W+D+U & 3 & 0.805 & 0.805 & 0.620 & 0.459 & 0.365 \\
W+D+X & 3 & 0.810 & 0.805 & 0.620 & 0.455 & 0.360 \\
W+D+U+X+M & 5 & 0.811 & 0.804 & 0.619 & 0.443 & 0.352 \\
W+U & 2 & 0.792 & 0.799 & 0.613 & 0.477 & 0.379 \\
W+U+X+M & 4 & 0.798 & 0.793 & 0.608 & 0.456 & 0.364 \\
W+X & 2 & 0.796 & 0.793 & 0.607 & 0.474 & 0.375 \\
U+X & 2 & 0.779 & 0.793 & 0.606 & 0.489 & 0.391 \\
D+U+X+M & 4 & 0.795 & 0.791 & 0.605 & 0.459 & 0.365 \\
D+X & 2 & 0.795 & 0.790 & 0.605 & 0.473 & 0.372 \\
W+D+U+M & 4 & 0.794 & 0.788 & 0.603 & 0.460 & 0.366 \\
D+U & 2 & 0.783 & 0.787 & 0.602 & 0.486 & 0.387 \\
W+D+X+M & 4 & 0.796 & 0.786 & 0.602 & 0.458 & 0.365 \\
W+U+M & 3 & 0.765 & 0.766 & 0.579 & 0.487 & 0.388 \\
U+X+M & 3 & 0.761 & 0.765 & 0.578 & 0.490 & 0.390 \\
W+X+M & 3 & 0.765 & 0.759 & 0.573 & 0.488 & 0.389 \\
D+U+M & 3 & 0.756 & 0.757 & 0.570 & 0.495 & 0.393 \\
D+X+M & 3 & 0.760 & 0.757 & 0.572 & 0.492 & 0.389 \\
W+D+M & 3 & 0.759 & 0.753 & 0.569 & 0.493 & 0.392 \\
U+M & 2 & 0.657 & 0.672 & 0.489 & 0.576 & 0.453 \\
W+M & 2 & 0.670 & 0.669 & 0.490 & 0.569 & 0.450 \\
X+M & 2 & 0.661 & 0.666 & 0.488 & 0.574 & 0.454 \\
D+M & 2 & 0.654 & 0.655 & 0.479 & 0.582 & 0.457 \\
\bottomrule\end{tabular}\end{minipage}\hfill
\begin{minipage}[t]{0.48\textwidth}\centering
\begin{tabular}{@{}l r rrrrr@{}}\toprule \multicolumn{7}{@{}c@{}}{\textbf{mos260}} \\ \cmidrule{1-7} Subset & $K$ & $r$ & $\rho$ & $\tau$ & RMSE & MAE \\ \midrule
D+U & 2 & 0.692 & 0.617 & 0.454 & 0.867 & 0.675 \\
D+U+X & 3 & 0.689 & 0.602 & 0.440 & 0.862 & 0.679 \\
W+D+U & 3 & 0.689 & 0.602 & 0.440 & 0.862 & 0.679 \\
D+X & 2 & 0.675 & 0.589 & 0.429 & 0.889 & 0.698 \\
W+D+U+X & 4 & 0.674 & 0.578 & 0.420 & 0.886 & 0.701 \\
W+D+X & 3 & 0.654 & 0.555 & 0.401 & 0.921 & 0.728 \\
W+U & 2 & 0.646 & 0.547 & 0.394 & 0.929 & 0.740 \\
U+X & 2 & 0.646 & 0.547 & 0.394 & 0.929 & 0.740 \\
W+U+X & 3 & 0.634 & 0.531 & 0.381 & 0.948 & 0.758 \\
W+D+U+X+M & 5 & 0.606 & 0.510 & 0.364 & 0.920 & 0.728 \\
D+U+X+M & 4 & 0.592 & 0.506 & 0.361 & 0.927 & 0.732 \\
W+D+U+M & 4 & 0.592 & 0.506 & 0.361 & 0.927 & 0.732 \\
D+U+M & 3 & 0.538 & 0.477 & 0.338 & 0.973 & 0.765 \\
W+X & 2 & 0.583 & 0.473 & 0.337 & 1.041 & 0.835 \\
W+D+X+M & 4 & 0.560 & 0.459 & 0.325 & 0.964 & 0.763 \\
W+U+X+M & 4 & 0.534 & 0.443 & 0.311 & 0.991 & 0.790 \\
D+X+M & 3 & 0.518 & 0.431 & 0.303 & 0.991 & 0.783 \\
W+D+M & 3 & 0.518 & 0.431 & 0.303 & 0.991 & 0.783 \\
W+U+M & 3 & 0.480 & 0.410 & 0.285 & 1.029 & 0.820 \\
U+X+M & 3 & 0.480 & 0.410 & 0.285 & 1.029 & 0.820 \\
W+X+M & 3 & 0.433 & 0.341 & 0.236 & 1.085 & 0.864 \\
D+M & 2 & 0.356 & 0.312 & 0.217 & 1.116 & 0.878 \\
U+M & 2 & 0.300 & 0.275 & 0.188 & 1.174 & 0.931 \\
W+M & 2 & 0.268 & 0.204 & 0.139 & 1.202 & 0.955 \\
X+M & 2 & 0.268 & 0.204 & 0.139 & 1.202 & 0.955 \\
\bottomrule\end{tabular}\end{minipage}
\end{table*}

\begin{table*}[t]\centering\small\setlength{\tabcolsep}{4pt}\renewcommand{\arraystretch}{0.93}
\caption{Teacher subsets under the $\rho$-weighted rule, 25 of 26, each side sorted by its own Spearman $\rho$.}
\label{tab:supp_spearman}
\begin{minipage}[t]{0.48\textwidth}\centering
\begin{tabular}{@{}l r rrrrr@{}}\toprule \multicolumn{7}{@{}c@{}}{\textbf{URGENT}} \\ \cmidrule{1-7} Subset & $K$ & $r$ & $\rho$ & $\tau$ & RMSE & MAE \\ \midrule
W+D+U+X+M & 5 & 0.816 & 0.814 & 0.630 & 0.442 & 0.351 \\
W+D+U+X & 4 & 0.813 & 0.814 & 0.629 & 0.448 & 0.356 \\
W+U+X+M & 4 & 0.809 & 0.809 & 0.625 & 0.450 & 0.358 \\
W+U+X & 3 & 0.804 & 0.809 & 0.623 & 0.459 & 0.365 \\
D+U+X+M & 4 & 0.807 & 0.806 & 0.622 & 0.452 & 0.359 \\
D+U+X & 3 & 0.802 & 0.806 & 0.621 & 0.461 & 0.366 \\
W+D+U+M & 4 & 0.809 & 0.805 & 0.621 & 0.450 & 0.358 \\
W+D+U & 3 & 0.806 & 0.805 & 0.620 & 0.459 & 0.364 \\
W+D+X & 3 & 0.810 & 0.805 & 0.620 & 0.455 & 0.360 \\
W+D+X+M & 4 & 0.813 & 0.805 & 0.621 & 0.447 & 0.353 \\
W+U & 2 & 0.793 & 0.799 & 0.613 & 0.476 & 0.379 \\
U+X+M & 3 & 0.787 & 0.795 & 0.609 & 0.473 & 0.377 \\
W+X & 2 & 0.796 & 0.793 & 0.607 & 0.474 & 0.375 \\
W+X+M & 3 & 0.801 & 0.793 & 0.608 & 0.461 & 0.365 \\
U+X & 2 & 0.779 & 0.793 & 0.606 & 0.490 & 0.391 \\
W+D & 2 & 0.799 & 0.791 & 0.606 & 0.471 & 0.372 \\
W+D+M & 3 & 0.802 & 0.790 & 0.606 & 0.460 & 0.363 \\
D+X+M & 3 & 0.799 & 0.790 & 0.606 & 0.461 & 0.363 \\
D+X & 2 & 0.795 & 0.790 & 0.605 & 0.473 & 0.372 \\
D+U+M & 3 & 0.789 & 0.788 & 0.603 & 0.472 & 0.375 \\
D+U & 2 & 0.783 & 0.787 & 0.602 & 0.486 & 0.387 \\
W+M & 2 & 0.784 & 0.770 & 0.585 & 0.480 & 0.380 \\
U+M & 2 & 0.737 & 0.750 & 0.561 & 0.526 & 0.418 \\
X+M & 2 & 0.755 & 0.747 & 0.563 & 0.509 & 0.401 \\
D+M & 2 & 0.764 & 0.746 & 0.564 & 0.501 & 0.392 \\
\bottomrule\end{tabular}\end{minipage}\hfill
\begin{minipage}[t]{0.48\textwidth}\centering
\begin{tabular}{@{}l r rrrrr@{}}\toprule \multicolumn{7}{@{}c@{}}{\textbf{mos260}} \\ \cmidrule{1-7} Subset & $K$ & $r$ & $\rho$ & $\tau$ & RMSE & MAE \\ \midrule
D+U+M & 3 & 0.694 & 0.619 & 0.456 & 0.866 & 0.673 \\
D+U & 2 & 0.694 & 0.619 & 0.456 & 0.866 & 0.673 \\
D+M & 2 & 0.670 & 0.613 & 0.456 & 0.927 & 0.712 \\
D+U+X+M & 4 & 0.693 & 0.608 & 0.446 & 0.858 & 0.673 \\
D+U+X & 3 & 0.693 & 0.608 & 0.446 & 0.858 & 0.673 \\
W+D+U+M & 4 & 0.693 & 0.608 & 0.446 & 0.858 & 0.673 \\
W+D+U & 3 & 0.693 & 0.608 & 0.446 & 0.858 & 0.673 \\
W+D & 2 & 0.680 & 0.598 & 0.438 & 0.882 & 0.691 \\
W+D+M & 3 & 0.680 & 0.598 & 0.438 & 0.882 & 0.691 \\
D+X+M & 3 & 0.680 & 0.598 & 0.438 & 0.882 & 0.691 \\
D+X & 2 & 0.680 & 0.598 & 0.438 & 0.882 & 0.691 \\
W+D+U+X+M & 5 & 0.680 & 0.588 & 0.429 & 0.875 & 0.691 \\
W+D+U+X & 4 & 0.680 & 0.588 & 0.429 & 0.875 & 0.691 \\
W+D+X & 3 & 0.663 & 0.568 & 0.413 & 0.907 & 0.716 \\
W+D+X+M & 4 & 0.663 & 0.568 & 0.413 & 0.907 & 0.716 \\
U+M & 2 & 0.622 & 0.555 & 0.403 & 0.991 & 0.781 \\
W+U & 2 & 0.648 & 0.550 & 0.397 & 0.927 & 0.738 \\
U+X+M & 3 & 0.648 & 0.550 & 0.397 & 0.927 & 0.738 \\
U+X & 2 & 0.648 & 0.550 & 0.397 & 0.927 & 0.738 \\
W+U+X+M & 4 & 0.638 & 0.535 & 0.385 & 0.942 & 0.752 \\
W+U+X & 3 & 0.638 & 0.535 & 0.385 & 0.942 & 0.752 \\
W+X & 2 & 0.583 & 0.473 & 0.337 & 1.041 & 0.835 \\
W+X+M & 3 & 0.583 & 0.473 & 0.337 & 1.041 & 0.835 \\
W+M & 2 & 0.583 & 0.473 & 0.337 & 1.041 & 0.835 \\
X+M & 2 & 0.583 & 0.473 & 0.337 & 1.041 & 0.835 \\
\bottomrule\end{tabular}\end{minipage}
\end{table*}

\begin{table*}[t]\centering\small\setlength{\tabcolsep}{4pt}\renewcommand{\arraystretch}{0.93}
\caption{Teacher subsets under the 1/RMSE-weighted rule, 25 of 26, each side sorted by its own Spearman $\rho$.}
\label{tab:supp_1rmse}
\begin{minipage}[t]{0.48\textwidth}\centering
\begin{tabular}{@{}l r rrrrr@{}}\toprule \multicolumn{7}{@{}c@{}}{\textbf{URGENT}} \\ \cmidrule{1-7} Subset & $K$ & $r$ & $\rho$ & $\tau$ & RMSE & MAE \\ \midrule
W+D+U+X & 4 & 0.813 & 0.814 & 0.629 & 0.448 & 0.355 \\
W+D+U+X+M & 5 & 0.818 & 0.811 & 0.627 & 0.436 & 0.346 \\
W+U+X & 3 & 0.805 & 0.809 & 0.623 & 0.459 & 0.365 \\
D+U+X & 3 & 0.803 & 0.806 & 0.621 & 0.460 & 0.366 \\
W+D+U & 3 & 0.806 & 0.805 & 0.621 & 0.458 & 0.364 \\
W+D+X & 3 & 0.810 & 0.805 & 0.620 & 0.455 & 0.360 \\
W+U+X+M & 4 & 0.810 & 0.804 & 0.619 & 0.444 & 0.353 \\
D+U+X+M & 4 & 0.807 & 0.801 & 0.617 & 0.447 & 0.355 \\
W+D+U+M & 4 & 0.808 & 0.800 & 0.615 & 0.446 & 0.355 \\
W+U & 2 & 0.794 & 0.799 & 0.613 & 0.475 & 0.378 \\
W+X & 2 & 0.796 & 0.794 & 0.607 & 0.474 & 0.375 \\
U+X & 2 & 0.779 & 0.793 & 0.606 & 0.489 & 0.391 \\
W+D & 2 & 0.799 & 0.791 & 0.606 & 0.471 & 0.372 \\
D+X & 2 & 0.795 & 0.790 & 0.605 & 0.473 & 0.372 \\
W+U+M & 3 & 0.793 & 0.788 & 0.601 & 0.461 & 0.368 \\
D+U & 2 & 0.784 & 0.787 & 0.602 & 0.486 & 0.386 \\
U+X+M & 3 & 0.785 & 0.784 & 0.598 & 0.469 & 0.373 \\
W+X+M & 3 & 0.793 & 0.782 & 0.597 & 0.461 & 0.368 \\
D+X+M & 3 & 0.789 & 0.778 & 0.594 & 0.465 & 0.368 \\
W+D+M & 3 & 0.791 & 0.778 & 0.593 & 0.464 & 0.369 \\
D+U+M & 3 & 0.782 & 0.776 & 0.591 & 0.472 & 0.375 \\
W+M & 2 & 0.743 & 0.732 & 0.547 & 0.509 & 0.406 \\
X+M & 2 & 0.722 & 0.716 & 0.533 & 0.526 & 0.418 \\
U+M & 2 & 0.711 & 0.715 & 0.527 & 0.535 & 0.424 \\
D+M & 2 & 0.720 & 0.711 & 0.530 & 0.530 & 0.417 \\
\bottomrule\end{tabular}\end{minipage}\hfill
\begin{minipage}[t]{0.48\textwidth}\centering
\begin{tabular}{@{}l r rrrrr@{}}\toprule \multicolumn{7}{@{}c@{}}{\textbf{mos260}} \\ \cmidrule{1-7} Subset & $K$ & $r$ & $\rho$ & $\tau$ & RMSE & MAE \\ \midrule
D+U & 2 & 0.693 & 0.619 & 0.455 & 0.866 & 0.674 \\
D+U+X & 3 & 0.691 & 0.605 & 0.443 & 0.860 & 0.676 \\
W+D+U & 3 & 0.691 & 0.605 & 0.443 & 0.860 & 0.676 \\
W+D & 2 & 0.678 & 0.593 & 0.433 & 0.886 & 0.695 \\
D+X & 2 & 0.678 & 0.593 & 0.433 & 0.886 & 0.695 \\
W+D+U+X & 4 & 0.677 & 0.583 & 0.424 & 0.881 & 0.697 \\
W+D+X & 3 & 0.658 & 0.561 & 0.406 & 0.914 & 0.723 \\
D+U+X+M & 4 & 0.647 & 0.557 & 0.403 & 0.880 & 0.691 \\
W+D+U+M & 4 & 0.647 & 0.557 & 0.403 & 0.880 & 0.691 \\
D+U+M & 3 & 0.628 & 0.554 & 0.400 & 0.896 & 0.701 \\
W+U & 2 & 0.647 & 0.548 & 0.395 & 0.928 & 0.739 \\
U+X & 2 & 0.647 & 0.548 & 0.395 & 0.928 & 0.739 \\
W+D+U+X+M & 5 & 0.644 & 0.547 & 0.394 & 0.889 & 0.703 \\
W+U+X & 3 & 0.635 & 0.532 & 0.382 & 0.946 & 0.756 \\
D+X+M & 3 & 0.608 & 0.513 & 0.368 & 0.917 & 0.721 \\
W+D+M & 3 & 0.608 & 0.513 & 0.368 & 0.917 & 0.721 \\
W+U+X+M & 4 & 0.585 & 0.486 & 0.344 & 0.956 & 0.762 \\
W+U+M & 3 & 0.564 & 0.477 & 0.337 & 0.965 & 0.768 \\
U+X+M & 3 & 0.564 & 0.477 & 0.337 & 0.965 & 0.768 \\
W+X & 2 & 0.583 & 0.473 & 0.337 & 1.041 & 0.835 \\
D+M & 2 & 0.534 & 0.464 & 0.331 & 0.980 & 0.765 \\
U+M & 2 & 0.451 & 0.416 & 0.289 & 1.062 & 0.842 \\
W+X+M & 3 & 0.504 & 0.400 & 0.280 & 1.040 & 0.828 \\
W+M & 2 & 0.402 & 0.316 & 0.218 & 1.106 & 0.880 \\
X+M & 2 & 0.402 & 0.316 & 0.218 & 1.106 & 0.880 \\
\bottomrule\end{tabular}\end{minipage}
\end{table*}

\begin{table*}[t]\centering\small\setlength{\tabcolsep}{4pt}\renewcommand{\arraystretch}{0.93}
\caption{Teacher subsets under the MSE-optimised rule, 25 of 26, each side sorted by its own Spearman $\rho$.}
\label{tab:supp_optimised}
\begin{minipage}[t]{0.48\textwidth}\centering
\begin{tabular}{@{}l r rrrrr@{}}\toprule \multicolumn{7}{@{}c@{}}{\textbf{URGENT}} \\ \cmidrule{1-7} Subset & $K$ & $r$ & $\rho$ & $\tau$ & RMSE & MAE \\ \midrule
W+D+U+X & 4 & 0.814 & 0.813 & 0.629 & 0.447 & 0.355 \\
W+D+U+X+M & 5 & 0.819 & 0.811 & 0.627 & 0.435 & 0.345 \\
W+U+X & 3 & 0.807 & 0.809 & 0.623 & 0.457 & 0.363 \\
W+U+X+M & 4 & 0.813 & 0.806 & 0.622 & 0.442 & 0.351 \\
W+D+U & 3 & 0.808 & 0.805 & 0.621 & 0.457 & 0.362 \\
W+D+X & 3 & 0.810 & 0.805 & 0.620 & 0.455 & 0.360 \\
D+U+X & 3 & 0.804 & 0.805 & 0.620 & 0.459 & 0.364 \\
D+U+X+M & 4 & 0.809 & 0.803 & 0.619 & 0.445 & 0.353 \\
W+D+U+M & 4 & 0.812 & 0.803 & 0.619 & 0.444 & 0.352 \\
W+D+X+M & 4 & 0.813 & 0.802 & 0.618 & 0.442 & 0.351 \\
W+U+M & 3 & 0.803 & 0.796 & 0.610 & 0.453 & 0.361 \\
W+X & 2 & 0.798 & 0.795 & 0.609 & 0.472 & 0.373 \\
W+D & 2 & 0.800 & 0.792 & 0.607 & 0.471 & 0.371 \\
U+X & 2 & 0.780 & 0.792 & 0.605 & 0.489 & 0.390 \\
W+X+M & 3 & 0.802 & 0.791 & 0.606 & 0.454 & 0.361 \\
U+X+M & 3 & 0.791 & 0.790 & 0.605 & 0.464 & 0.369 \\
D+X & 2 & 0.795 & 0.790 & 0.605 & 0.473 & 0.372 \\
W+D+M & 3 & 0.802 & 0.788 & 0.604 & 0.456 & 0.361 \\
D+X+M & 3 & 0.799 & 0.788 & 0.603 & 0.457 & 0.360 \\
D+U & 2 & 0.785 & 0.786 & 0.600 & 0.484 & 0.385 \\
D+U+M & 3 & 0.791 & 0.784 & 0.599 & 0.465 & 0.369 \\
W+M & 2 & 0.783 & 0.767 & 0.582 & 0.478 & 0.380 \\
D+M & 2 & 0.761 & 0.744 & 0.562 & 0.500 & 0.392 \\
X+M & 2 & 0.754 & 0.743 & 0.559 & 0.503 & 0.397 \\
U+M & 2 & 0.739 & 0.741 & 0.552 & 0.515 & 0.410 \\
\bottomrule\end{tabular}\end{minipage}\hfill
\begin{minipage}[t]{0.48\textwidth}\centering
\begin{tabular}{@{}l r rrrrr@{}}\toprule \multicolumn{7}{@{}c@{}}{\textbf{mos260}} \\ \cmidrule{1-7} Subset & $K$ & $r$ & $\rho$ & $\tau$ & RMSE & MAE \\ \midrule
D+U & 2 & 0.692 & 0.617 & 0.454 & 0.867 & 0.675 \\
W+D+U & 3 & 0.689 & 0.602 & 0.440 & 0.862 & 0.679 \\
D+U+X & 3 & 0.689 & 0.602 & 0.440 & 0.862 & 0.679 \\
W+D & 2 & 0.675 & 0.589 & 0.429 & 0.889 & 0.698 \\
D+X & 2 & 0.675 & 0.589 & 0.429 & 0.889 & 0.698 \\
W+D+U+X & 4 & 0.674 & 0.578 & 0.420 & 0.886 & 0.701 \\
W+D+X & 3 & 0.654 & 0.555 & 0.401 & 0.921 & 0.728 \\
U+X & 2 & 0.646 & 0.547 & 0.394 & 0.929 & 0.740 \\
W+U+X & 3 & 0.634 & 0.531 & 0.381 & 0.948 & 0.758 \\
W+D+U+X+M & 5 & 0.606 & 0.510 & 0.364 & 0.920 & 0.728 \\
D+U+X+M & 4 & 0.592 & 0.506 & 0.361 & 0.927 & 0.732 \\
W+D+U+M & 4 & 0.592 & 0.506 & 0.361 & 0.927 & 0.732 \\
D+U+M & 3 & 0.538 & 0.477 & 0.338 & 0.973 & 0.765 \\
W+X & 2 & 0.583 & 0.473 & 0.337 & 1.041 & 0.835 \\
W+D+X+M & 4 & 0.560 & 0.459 & 0.325 & 0.964 & 0.763 \\
W+U+X+M & 4 & 0.534 & 0.443 & 0.311 & 0.991 & 0.790 \\
W+D+M & 3 & 0.518 & 0.431 & 0.303 & 0.991 & 0.783 \\
D+X+M & 3 & 0.518 & 0.431 & 0.303 & 0.991 & 0.783 \\
W+U+M & 3 & 0.480 & 0.410 & 0.285 & 1.029 & 0.820 \\
U+X+M & 3 & 0.480 & 0.410 & 0.285 & 1.029 & 0.820 \\
W+X+M & 3 & 0.433 & 0.341 & 0.236 & 1.085 & 0.864 \\
D+M & 2 & 0.356 & 0.312 & 0.217 & 1.116 & 0.878 \\
U+M & 2 & 0.300 & 0.275 & 0.188 & 1.174 & 0.931 \\
W+M & 2 & 0.268 & 0.204 & 0.139 & 1.202 & 0.955 \\
X+M & 2 & 0.268 & 0.204 & 0.139 & 1.202 & 0.955 \\
\bottomrule\end{tabular}\end{minipage}
\end{table*}

\begin{table*}[t]\centering\small\setlength{\tabcolsep}{4pt}\renewcommand{\arraystretch}{0.93}
\caption{Teacher subsets under the adaptive router rule, 25 of 26, each side sorted by its own Spearman $\rho$.}
\label{tab:supp_router}
\begin{minipage}[t]{0.48\textwidth}\centering
\begin{tabular}{@{}l r rrrrr@{}}\toprule \multicolumn{7}{@{}c@{}}{\textbf{URGENT}} \\ \cmidrule{1-7} Subset & $K$ & $r$ & $\rho$ & $\tau$ & RMSE & MAE \\ \midrule
W+D+U+X+M & 5 & 0.825 & 0.817 & 0.633 & 0.428 & 0.340 \\
W+D+U+X & 4 & 0.819 & 0.816 & 0.632 & 0.439 & 0.348 \\
W+U+X & 3 & 0.809 & 0.810 & 0.624 & 0.451 & 0.358 \\
W+U+X+M & 4 & 0.818 & 0.810 & 0.625 & 0.436 & 0.346 \\
W+D+U+M & 4 & 0.816 & 0.807 & 0.623 & 0.438 & 0.348 \\
D+U+X & 3 & 0.808 & 0.806 & 0.622 & 0.451 & 0.357 \\
W+D+U & 3 & 0.809 & 0.806 & 0.621 & 0.452 & 0.358 \\
W+D+X & 3 & 0.811 & 0.806 & 0.621 & 0.452 & 0.357 \\
D+U+X+M & 4 & 0.814 & 0.805 & 0.622 & 0.440 & 0.348 \\
W+D+X+M & 4 & 0.815 & 0.805 & 0.621 & 0.438 & 0.348 \\
W+U+M & 3 & 0.807 & 0.797 & 0.613 & 0.447 & 0.356 \\
W+X & 2 & 0.798 & 0.795 & 0.609 & 0.472 & 0.373 \\
W+X+M & 3 & 0.805 & 0.794 & 0.609 & 0.451 & 0.358 \\
U+X & 2 & 0.785 & 0.793 & 0.606 & 0.479 & 0.381 \\
W+D & 2 & 0.800 & 0.792 & 0.607 & 0.470 & 0.371 \\
U+X+M & 3 & 0.797 & 0.792 & 0.606 & 0.458 & 0.363 \\
D+X & 2 & 0.795 & 0.790 & 0.605 & 0.472 & 0.371 \\
W+D+M & 3 & 0.803 & 0.789 & 0.605 & 0.453 & 0.359 \\
D+X+M & 3 & 0.801 & 0.789 & 0.604 & 0.454 & 0.358 \\
D+U & 2 & 0.789 & 0.788 & 0.603 & 0.478 & 0.380 \\
D+U+M & 3 & 0.796 & 0.786 & 0.601 & 0.460 & 0.363 \\
W+M & 2 & 0.782 & 0.769 & 0.584 & 0.480 & 0.381 \\
X+M & 2 & 0.754 & 0.745 & 0.561 & 0.501 & 0.396 \\
D+M & 2 & 0.758 & 0.744 & 0.561 & 0.501 & 0.393 \\
U+M & 2 & 0.737 & 0.737 & 0.548 & 0.514 & 0.409 \\
\bottomrule\end{tabular}\end{minipage}\hfill
\begin{minipage}[t]{0.48\textwidth}\centering
\begin{tabular}{@{}l r rrrrr@{}}\toprule \multicolumn{7}{@{}c@{}}{\textbf{mos260}} \\ \cmidrule{1-7} Subset & $K$ & $r$ & $\rho$ & $\tau$ & RMSE & MAE \\ \midrule
W+D+U+X+M & 5 & 0.708 & 0.636 & 0.471 & 0.812 & 0.625 \\
W+D+U+X & 4 & 0.706 & 0.636 & 0.471 & 0.822 & 0.633 \\
D+U+X & 3 & 0.705 & 0.635 & 0.469 & 0.824 & 0.634 \\
D+U & 2 & 0.703 & 0.635 & 0.471 & 0.837 & 0.641 \\
D+U+X+M & 4 & 0.706 & 0.633 & 0.468 & 0.819 & 0.630 \\
W+D+U+M & 4 & 0.706 & 0.632 & 0.468 & 0.815 & 0.626 \\
W+D+U & 3 & 0.703 & 0.631 & 0.466 & 0.823 & 0.636 \\
D+U+M & 3 & 0.697 & 0.625 & 0.462 & 0.826 & 0.635 \\
W+D & 2 & 0.686 & 0.622 & 0.458 & 0.867 & 0.672 \\
D+X & 2 & 0.690 & 0.622 & 0.459 & 0.870 & 0.674 \\
W+D+X+M & 4 & 0.685 & 0.619 & 0.456 & 0.844 & 0.651 \\
D+X+M & 3 & 0.686 & 0.619 & 0.455 & 0.844 & 0.652 \\
W+D+X & 3 & 0.680 & 0.614 & 0.451 & 0.866 & 0.674 \\
W+D+M & 3 & 0.681 & 0.612 & 0.449 & 0.845 & 0.653 \\
D+M & 2 & 0.652 & 0.604 & 0.442 & 0.886 & 0.679 \\
U+X & 2 & 0.652 & 0.554 & 0.401 & 0.896 & 0.709 \\
W+U+X & 3 & 0.651 & 0.550 & 0.397 & 0.894 & 0.710 \\
W+U+M & 3 & 0.657 & 0.543 & 0.392 & 0.881 & 0.695 \\
W+U+X+M & 4 & 0.652 & 0.542 & 0.391 & 0.883 & 0.695 \\
U+X+M & 3 & 0.654 & 0.539 & 0.388 & 0.888 & 0.701 \\
U+M & 2 & 0.582 & 0.505 & 0.357 & 0.945 & 0.742 \\
W+X & 2 & 0.583 & 0.473 & 0.337 & 1.041 & 0.835 \\
W+X+M & 3 & 0.556 & 0.449 & 0.317 & 0.980 & 0.780 \\
W+M & 2 & 0.550 & 0.447 & 0.316 & 1.003 & 0.800 \\
X+M & 2 & 0.550 & 0.447 & 0.316 & 1.003 & 0.800 \\
\bottomrule\end{tabular}\end{minipage}
\end{table*}